\documentclass[aps]{revtex4}
    \def\be{\begin{equation}}
    \def\ee{\end{equation}}
    \def\ba{\begin{eqnarray}}
    \def\ea{\end{eqnarray}}

\begin{document}
\title{Back-reaction of Cosmological Fluctuations during Power-Law Inflation}

\author{G. Marozzi} \email{marozzi@bo.infn.it}

\affiliation{Dipartimento di Fisica, Universit\`a degli Studi di Bologna
    and I.N.F.N., \\ via Irnerio, 46 -- I-40126 Bologna -- Italy}


\begin{abstract}
We study the renormalized energy-momentum tensor of
cosmological scalar fluctuations during the slow-rollover regime for
power-law inflation and find that it is characterized by a negative
energy density at the leading order, with the same time behaviour as the
background energy.
The average expansion rate appears decreased by the back-reaction of the
effective energy  of
cosmological fluctuations, but this value is comparable with the energy
of background only if inflation starts at a Planckian energy.
We also find that, for this particular model, the first and second order
inflaton fluctuations are decoupled and satisfy the same equation of motion.
To conclude, the fourth order adiabatic
expansion for the inflaton scalar field is evaluated for a general potential
$V(\phi)$.
\end{abstract}

\maketitle


\section{Introduction}\label{one}

Inflationary models (see  \cite{lindebook,kolbturner} for a textbook
review), born to solve the problems associated with the standard big-bang
theory, have their greatest success in the generation of  the initial
condition for the large scale structure of the Universe through the
amplification
of the quantum cosmological fluctuations during their accelerated era
\cite{all}.
Using the last cosmic microwave background (CMB) data (\cite{wmap06}) 
and considering the
transition from inflation to the standard Big Bang cosmology the constraints
on the amplitude and spectrum of CMB fluctuation become constraints on the
different models of inflation.
All these constraints depend on the linear treatment of cosmological
perturbations, so, on using the most recent data, we are close to measuring
the details of the spectrum of fluctuations.
What is still not completely understood is their energy content
and the eventual back-reaction on the inflationary expansion
responsible for their amplification.

Among the interesting effects, appreciable only beyond the linear order, one
notes, for example, the effect of the non-Gaussianities in the matter power
spectrum and in CMB anisotropies (see \cite{BKMR_report} for a review).
However, from the theoretical
point of view, we feel that the back-reaction of gravitational fluctuations
on the geometry
is one of the most interesting issues \cite{tsamis}.
Within the inflationary context, this problem has been tackled in
\cite{tsamis, ABML, ABMP}
with the intringuing result that the energy-momentum tensor (EMT henceforth)
of fluctuations may slow down inflation. Thanks to this result
in the last decade there has been
a renewed interest in the subject of back-reaction \cite{UNRUH, AW, AW2, GB,
FMVV69, BL, MB, LU}.

The aim of this paper is to compute the renormalized EMT of
cosmological fluctuations for a power-law model of inflation,
according to the adiabatic regularization scheme \cite{birrell} already used
in previous papers \cite{FMVV65,FMVV69}.
While the adiabatic vacuum for a cosmological scalar fluctuations, and
its associated EMT, can
be computed for generic space-times, the unrenormalized EMT
can be calculated analytically only if exact analytic solutions for the
field Fourier modes are available.
This is the case for the power-law inflation \cite{exponential}
where analytic solutions for a scalar field are available.
The main theoretical problem which arises is that inflation never comes to an
end because of its slow-roll parameter $\epsilon$ is constant.

Power-law solutions can be also applied in different backgrounds as,
for example,
the alternative cosmological scenario called Ekpyrotic Universe \cite{KOST},
and in the Pre-Big Bang scenario (for a review see \cite{VG}) in
string cosmology.

The plan of the paper is as follow. In Sec. \ref{two}  we present the linear
cosmological perturbations in the uniform curvature gauge (UCG henceforth).
In Sec. \ref{three} we extend the UCG to second order and using the Einstein
equations to second order we give the equation of motion for the second order
inflaton fluctuation.
In Sec. \ref{four} we
give the solutions for first order fluctuations for the particular case of
power-law inflation. In Sec. \ref{five} we compute the renormalized value
for the correlator.
We discuss the EMT and the back-reaction on the geometry in Sec. \ref{six}
and we give our
conclusions in Sec. \ref{seven}. In the Appendix we
exhibit the fourth order adiabatic expansion for a general potential $V(\phi)$.


\section{Linear Perturbations in the Uniform Curvature Gauge} \label{two}
We consider inflation in a flat universe driven by a classical minimally
coupled scalar field with a general potential $V(\phi)$.
The action is:
    \be
    S \equiv \int d^4x {\cal L} = \int d^4x \sqrt{-g} \left[
\frac{R}{16{\pi}G}
    - \frac{1}{2} g^{\mu \nu}
    \partial_{\mu} \phi \partial_{\nu} \phi
    - V(\phi) \right] \,,
    \label{action}
\ee
where ${\cal L}$ is the Lagrangian density.

Let us study the fluctuations of the scalar field $\varphi (t,{\bf
x})$ around its homogeneous classical
value $\phi (t)$ and
include metric perturbations. For the homogeneous case we have
\begin{eqnarray}
&& \ddot{\phi}+3H\dot{\phi}+V_\phi=0 \,, \label{eqmotion1}  \\
&& H^2 = \frac{1}{3 M _{pl}^2} \left[ \frac{\dot \phi^2}{2} + V \right]\,,
\label{EE00}
\end{eqnarray}
where $H=\dot a /a$ is the Hubble parameter and $a$ is the scale factor,
using the notation
$ M_{\rm pl}^2=1/(8\pi G)$ for the (reduced) Planck mass definition.

The scalar perturbations around a spatially flat Robertson-Walker metric are
 ($ds^2 = g_{\mu \nu} d x^\mu d x^\nu$):
\begin{eqnarray}
g_{00} &=& - 1 - 2 \alpha \nonumber \\
g_{0i} &=& - \frac{a}{2} \beta_{,i}
\nonumber \\
g_{ij} &=& a^2 \left[ \delta_{ij} (1 -2 \psi)
 + D_{i j} E \right] \,,
\label{metric_first}
\end{eqnarray}
where $D_{ij}=\partial_i \partial_j -1/3 \, \nabla^2 \delta_{ij}$.
In the above we neglect vector and tensor perturbations. If we
choose to work in the UCG one obtains: \be ds^2=-(1+2 \alpha)dt^2
- a \beta_{,i} dt dx^i+a^2 \delta_{ij} dx^i dx^j \,.
\label{PER_UCG2} \ee We note that this choice fixes uniquely the
gauge, just as the more frequently used longitudinal gauge (for a
review of cosmological perturbations in this gauge see
\cite{MFB}). The equation of motion for the first order scalar
field fluctuations in the UCG is given by \cite{FMVV69}: \be
\ddot{\varphi}+3 H \dot{\varphi}-\frac{1}{a^2}\nabla^2 \varphi +
V_{\phi \phi} \varphi = \dot{\alpha}\dot{\phi} - 2 \alpha V_\phi
-\frac{\dot{\phi}}{2a} \nabla^2\beta \,, \label{EQMP4} \ee where a
dot denotes a derivative w.r.t. the time $t$. Using the energy and
momentum constraints in their linearized version
\begin{eqnarray}
\frac{H}{a} \nabla^2 \beta &=& \frac{1}{M _{pl}^2} \left ( \dot{\phi}
\dot{\varphi} + V_\phi \varphi + 2 \, V \alpha \right)
\nonumber \\
&=& \frac{1}{M _{pl}^2} \frac{\dot{\phi}^2}{H} \, \frac{d}{dt} \left
(\frac{H}{\dot{\phi}}\varphi \right )\,,
\label{Eq_beta} \\
\alpha_{,i} &=& \frac{1}{2 M _{pl}^2} \frac{\dot{\phi}}{H} \, \varphi_{,i}
\label{Eq_alpha}
\end{eqnarray}
one obtains
\be
\ddot{\varphi} + 3 H \dot{\varphi} - \frac{1}{a^2}\nabla^2 \varphi +
\left[ V_{\phi \phi} + 2 \left(3 H +
\frac{\dot H}{H}\right)^.\right] \varphi = 0 \,.
\label{Eq_mukhanov}
\ee
The Fourier tranform modes of $v = a \varphi$ satisfy:
\be
v_k'' + \left( k^2 - \frac{z''}{z} \right) v_k = 0 \,, \quad z = a
\frac{\dot \phi}{H} \,,
\label{Eq_resc_mukhanov}
\ee
where $'$ denotes a derivative with respect to the conformal time $\eta$,
$d \eta = d t/a$.
On comparing the last equation with Eq. (12) of \cite{mukhanov}
it is immediate to see
that $\varphi$ satisfies the
same equation as the Mukhanov variable $Q$. Therefore,
the uniform curvature gauge has the advantage of singling out the
true dynamical degrees of freedom (the matter ones), even if it has the
disadvantage of being non diagonal in the metric perturbations.


\section{Beyond the Linear Order} \label{three}

To second order we consider a metric having the following coefficients:
\begin{eqnarray}
g_{00} &=& - 1 - 2 \alpha - 2 \alpha^{(2)} \nonumber \\
g_{0i} &=& - \frac{a}{2} \left( \beta_{,i} + \beta_{,i}^{(2)} \right)
\nonumber \\
g_{ij} &=& a^2 \left[ \delta_{ij} + \frac{1}{2} \left( \partial_i
\chi_{j}^{(2)} +
\partial_j \chi_{i}^{(2)} + h_{ij}^{(2)} \right) \right] \,.
\label{metric_second}
\end{eqnarray}
The above metric is the extension of the uniform curvature gauge to second
order: $\alpha^{(2)}$ and $\beta^{(2)}$ are the scalar perturbations to
second order. To second order, scalar, vector and tensor perturbations do
not evolve independently as is the case in first order. For this reason we
take into account second order vector and tensor perturbations,
represented by
the divergenceless vector $\chi_j^{(2)}$ and by the transverse and
traceless tensor $h_{ij}^{(2)}$, respectively. In the above we have
omitted first order vector perturbations (which die away kinematically)
and tensor perturbations (which satisfy the usual equation $\ddot h + 3 H
\dot h - \nabla^2 h /a^2 = 0$).
With this approximation we are neglecting the EMT of
vector and tensor perturbations, and their correlations with the scalar
perturbations.
We finally note that the choice in Eq. (\ref{metric_second}) (including
vector and tensor metric elements to first order) fixes the
gauge completely to second order.

Using the energy and momentum contrains up to second order we can obtain
the equations of motion for $\alpha^{(2)}$, $\beta^{(2)}$ and $\varphi^{(2)}$,
in particular one obtains for the second order field fluctuation the following
equation (see \cite{FMVV69})

\begin{eqnarray}
& & \ddot{\varphi}^{(2)} + 3 H \dot{\varphi}^{(2)} - \frac{1}{a^2}\nabla^2
\varphi^{(2)} + \left[ V_{\phi \phi} + 2 \left(3 H +
\frac{\dot H}{H}\right )^. \right] \varphi^{(2)} =
-\frac{1}{2} V_{\phi\phi\phi} \varphi^2+\frac{1}{M _{pl}^2}
\frac{\dot{\phi}}{2 H}\left[\left(-\frac{5}{2} V_{\phi\phi}-9 \dot{H} \right.
\right.
    \nonumber
\\ & & \left. \left.
+\frac{\dot{H}^2}{H^2}-2 \frac{\ddot{H}}{H}\right)\varphi^2-\frac{1}{2}
\dot{\varphi}^2+\left( -2 \frac{\dot H}{H}+ \frac{\ddot{H}}{\dot{H}}\right)
\varphi\dot{\varphi}+\frac{2}{a^2}\varphi \nabla^2 \varphi+\frac{1}{2 a^2}
|\vec{\nabla} \varphi|^2
\right]-\frac{\dot{\phi}}{16 H}\frac{1}{a^2}\left[ \beta_{,ij} \, \beta^{,ij}
-\left( \nabla^2 \beta \right)^2 \right]  \nonumber
\\ & & -\frac{H}{a} \vec{\nabla}\beta \cdot \vec{\nabla}\varphi-\frac{1}{2 a}
\vec{\nabla}\dot{\beta} \cdot \vec{\nabla}\varphi-\frac{1}{a}
\vec{\nabla}\beta \cdot \vec{\nabla}\dot{\varphi}-\dot{\phi}
\left(-3H+\frac{\dot H}{H}-\frac{\ddot{H}}{\dot{H}}\right)\tilde{s}+\dot{\phi}
\dot{\tilde{s}}
\label{Eq_mukhanov_second}
\end{eqnarray}
where $\tilde{s}$ is a  non-local spatial contribution given by
\be \tilde{s}=\frac{1}{\nabla^2} \left[\frac{1}{2 M _{pl}^2 H}
\vec{\nabla} \cdot \left( \dot \varphi \vec{\nabla} \varphi
\right) + \frac{1}{a}\frac{1}{M _{pl}^2}\frac{\dot{\phi}}{8 H^2}
\left( \varphi^{,kj} \beta_{,kj} - \varphi^{,k}_{,k}
\beta^{,j}_{,j} \right) \right] \,. \label{nonlocal} \ee


\section{Power-Law Solution for Linear Perturbation} \label{four}
We now want restrict ourselves to the case of power-law inflation
\cite{exponential} where $a(t)\sim t^p$ with $p>1$.
In this case we have an exponential potential
\be
V(\phi)=V_0 \exp \left[-\frac{\lambda}{M_{pl}}(\phi-\phi_i)\right]
\ee
with $V_0=\frac{M_{pl}^2}{t_i^2} p(3p-1)$ and
$\lambda=\left(\frac{2}{p}\right)^{1/2}$.
The scale factor and the omogeneus solution are given by
$$
a(t)=\left(\frac{t}{t_i}\right)^p \,,
$$
$$
\phi(t)=\phi_i+M_{pl}(2 p)^{1/2} \log \frac{t}{t_i}
$$
respectively, and the slow-roll parameters by
$$
\epsilon \equiv \frac{M _{pl}^2}{2}\left(\frac{V_\phi}{V}\right)^2=\frac{1}{p}
\,,\,\,\,\,\,\,\,\,\,\,\,\,\,\,\,\,\,\,\,\,\,\,\,\,\,\,\,
\eta \equiv M_{pl}^2 \frac{V_{\phi\phi}}{V}=\frac{2}{p}\,.
$$
The equation of motion (\ref{Eq_mukhanov}) has an exact solution
for this particular case \cite{exponential}.
In Fourier space, substituting the effective mass squared
$V_{\phi \phi}=\frac{2}{t^2}(3p-1)$
and the Hubble factor $H=p/t$, one obtains
\be
\ddot{\varphi}_k+3 H \dot{\varphi}_k+
\frac{k^2}{a^2} \varphi_k=0 \,.
\label{EQMPF_k}
\ee
Introducing $\psi_k=a^{3/2} \varphi_k$ and the conformal time $\eta$ the
equation becomes
\be
\psi_k''-\frac{p}{1-p}\frac{1}{\eta} \psi_k'+
\left\{k^2+\left[-\frac{9}{4} p^2+ \frac{3}{2} p\right]\frac{1}{(1-p)^2}
\frac{1}{\eta^2}\right\}\psi_k=0 \,
\label{EQMPF_k2}
\ee
which gives the general solution
\be
    \varphi_k = \frac{1}{a^{3/2}}
\left( -k\eta\right)^{\frac{1}{2}\frac{1}{1-p}} \left[ A H_\nu^{(1)}
    (-k \eta) + B H_\nu^{(2)} (-k \eta) \right]
    \label{solution}
\ee
with $\nu=\frac{3}{2}+\frac{1}{p-1}$.
We now consider quantized fluctuations of the inflaton:
\be
    \hat{\varphi} (\eta, {\bf x}) = \frac{1}{(2 \pi)^{3/2}}  
    \int  d {\bf k}
    \left[ \varphi_{k} (\eta) \, e^{i {\bf k} \cdot {\bf x}} \, \hat{b}_{\bf
    k} + \varphi^*_{k} (\eta) e^{- i {\bf k} \cdot {\bf x}}
    \hat{b}^\dagger_{{\bf k}} \right]
    \label{quantumFourier}
    \ee
where the $\hat{b}_k$ are time-independent Heisenberg operators.
In order to have the usual commutation relations among the $\hat{b}_k$:
 \be
    [\hat{b}_{\bf k}, \hat{b}_{{\bf k}'}] =
    [\hat{b}^\dagger_{\bf k}, \hat{b}^\dagger_{{\bf k}'}] = 0 \, \quad
    [\hat{b}_{\bf k}, \hat{b}^{\dagger}_{{\bf k}'}] = \delta^{(3)}
    ({\bf k} - {\bf k}') \,
\ee
one must normalize the solution to the equations of
motion through the Wronskian condition:
\be
    \varphi_k \, {\varphi'}^{*}_k -
    {\varphi}_k' \varphi^*_k = \frac{i}{a^2} \,.
    \label{wronskian}
\ee
This normalization condition yields the following
relation among the coefficients $A, B$ of Eq. (\ref{solution}):
\be
    |A|^2 - |B|^2 = -(-k \eta)^{\frac{1}{p-1}} a \frac{\pi \eta}{4} \,.
\ee
The solution associated with the adiabatic vacuum for $k \rightarrow \infty$
corresponds to choosing $A=(-k \eta)^{\frac{1}{2(p-1)}} a^{1/2}
\left(-\frac{\pi \eta}{4}\right)^{1/2}$ and $B=0$.
Thus one obtains the following adiabatic solution:
\be
\varphi_k=\frac{1}{a} \left(-\frac{\pi \eta}{4}\right)^{1/2} H_\nu^{(1)}
    (-k \eta)
\ee
which, in the proper time, becomes
\be
\varphi_k=\frac{1}{a^{3/2}} \left(\frac{\pi}{4 H}\right)^{1/2}
\left(\frac{p}{p-1}\right)^{1/2}
H_\nu^{(1)}\left(\frac{p}{p-1}\frac{k}{a H}\right)\,.
\label{solutionpropertime}
\ee


\section{Adiabatic Subtraction} \label{five}
We now compute the integrals obtained on taking the vacuum expectation values
of the relevant operators. Let us consider the correlator
$\langle\varphi^2\rangle$ (we will link this to the energy-momentum tensor in
the next section)
\be
\langle \varphi^2 \rangle =\frac{1}{(2 \pi)^3} \int_{|k|>l} d {\bf k} \,
|\varphi_k|^2
\label{fisquare}
\ee
where $\ell=C H_i$ is an infrared cut-off related to the beginning of
inflation \cite{VF,cutoff}.
Using solution (\ref{solutionpropertime}) we obtain a UV divergence, so we
employ, as in the previous papers \cite{FMVV65, FMVV69}, dimensional
regularization to treat the UV behaviour,
therefore the integrands will be in $3$ dimensions and the integration measure
analytically continued in $d$ dimensions.
Subsequently an adiabatic subtraction is performed in order to obtain the
renormalized quantities.
The $\langle\varphi^2\rangle$ becomes
\be
\langle \varphi^2 \rangle =\frac{1}{(2 \pi)^d} \frac{2 \pi^{d/2}}{\Gamma(d/2)}
\int_{\ell}^{+\infty} d k \, k^{d-1} \frac{\pi}{4 H}\frac{1}{a^3}
\frac{p}{p-1} \left[ J_\nu \left( \frac{p}{p-1}\frac{k}{a H}\right)^2+
N_\nu \left( \frac{p}{p-1}\frac{k}{a H}\right)^2 \right]
\label{fisquareDdimendions}
\ee
On using (B.3) of Appendix B of  \cite{FMVV65}   with $\alpha=d-1$ one obtains:
\begin{eqnarray}
\langle \varphi^2\rangle &=& \frac{1}{16\pi^2}
    H^2 \left\{ 2 p
\left(\frac{H_i}{H}\right)^2 \left(\frac{\Gamma(\nu)}{\Gamma(3/2)}\right)^2
\left(1-\frac{1}{p}\right)^{2 \nu} \left(\frac{\ell}{2 H_i}
\right)^{-\frac{2}{p-1}}
+\frac{2 p-1}{p} \pi \cot \left(-\frac{\pi}{p-1}\right)
+\left(\frac{1}{2 \pi^{1/2}} \right)^{d-3}
\left(\frac{p-1}{p} \right)^{d-3} \right.
\nonumber \\  & & \left.
\left( a H \right)^{d-3}
\frac{\Gamma(3/2)}{\Gamma(d/2)}\, \frac{\Gamma(-1/2)
\Gamma(-1/2+\nu)}{\Gamma(3/2+\nu)}\,
\frac{\Gamma(d/2+\nu)}
{\Gamma(1-d/2) \Gamma(1-d/2+\nu)} \left(-2+\frac{1}{p}\right)
\Gamma \left(\frac{1}{2}-\frac{d}{2}\right)
+{\cal O} \left(\left(\frac{1}{a}\right)^{2-\frac{2}{p-1}} \right)\right\}
\nonumber \\ & &
+{\cal O} (d-3)
\label{phi2NR}
 \end{eqnarray}
where to avoid singularities in the analytic continuation we have to consider
$p>3$, this is not a problem because, for power-law inflation,
the range $p<60$ is disfavored at 2 $\sigma$
(see, for example, \cite{ObsPL}).

The adiabatic fourth order is (see Appendix and \cite{FMVV65}):
\begin{eqnarray}
\langle \varphi^2\rangle_{(4)} &=& \frac{1}{16\pi^2}
    H^2 \left\{ \frac{17}{10} \frac{1}{p^2}-\frac{131}{30} \frac{1}{p}+
\frac{31}{60}+\frac{5}{12} p-\frac{17}{60}\frac{p^2}{3 p-1}
+
\left(\frac{1}{2 \pi^{1/2}} \right)^{d-3}
\left( a V_{\phi\phi} \right)^{d-3}
\right.
\nonumber \\  & & \left.
\left(-2+\frac{1}{p}\right)
\Gamma \left(\frac{1}{2}-\frac{d}{2}\right)
+{\cal O} \left(\frac{1}{a^3} \right)\right\}
+{\cal O} (d-3)
\label{phi2AD}
 \end{eqnarray}
and the resulting renormalized quantity is
\begin{eqnarray}
\langle\varphi^2\rangle_{\rm REN} &=&\lim_{d\rightarrow 3}
\left(\langle\varphi^2\rangle
-\langle\varphi^2\rangle_{(4)} \right) \nonumber \\
&=&
\frac{1}{16\pi^2}
    H^2 \left\{2 p
\left(\frac{H_i}{H}\right)^2 \left(\frac{\Gamma(\nu)}{\Gamma(3/2)}\right)^2
\left(1-\frac{1}{p}\right)^{2 \nu} \left(\frac{\ell}{2 H_i}
\right)^{-\frac{2}{p-1}}+
\frac{2 p-1}{p} \pi \cot \left(-\frac{\pi}{p-1}\right)-
\frac{17}{10} \frac{1}{p^2}+\frac{131}{30} \frac{1}{p}-
\frac{31}{60}\right.
\nonumber \\  & & \left.
-\frac{5}{12} p+\frac{17}{60}\frac{p^2}{3 p-1}
+\left[\log\left(\frac{p-1}{2(3 p-1)^{1/2}} \right)+\frac{1}{2}
\left(\frac{p-1}{2 p-1} +\frac{p-1}{p}+2 \Psi\left(1+
\frac{1}{p-1}\right) \right) \right] \right.
\nonumber \\ & & \left.
\left(-4+\frac{2}{p}\right)+
{\cal O} \left(\left(\frac{1}{a}\right)^{2-\frac{2}{p-1}} \right)\right\}\,.
\label{phi2RIN}
\end{eqnarray}

So the leading behaviour of the renormalized correlator (which is essentially
given by the infrared contribution of $\langle\varphi^2\rangle$, namely the
first term of Eq.(\ref{phi2NR})), during the inflationary era, with $p>>1$, is:
\be
\langle\varphi^2\rangle_{\rm REN} \sim \frac{p}{8 \pi^2} H_i^2 \,,
\label{phiRENleading}
\ee
the same result is obtained in \cite{VSahni} for a massless minimally coupled
test scalar field in a power-law model of inflation. In fact this test scalar
field satisfies the same equation of motion of the field fluctuation
$\varphi$ (see Eq.(\ref{EQMPF_k})).

Let us comment on the possible gauge dependence of this result. Starting
from the general metric (\ref{metric_first}) and considering the
first order gauge transformation (see, for example, \cite{SOGIP})
the gauge invariant Mukhanov variable $Q$ is given by
$\varphi+\frac{\dot{\phi}}{H} \psi$, and in the UCG this gauge
invariant variable, as already stated, coincides with the scalar
field fluctuation $\varphi$. Thus, in this particular gauge,
$\varphi^2$ coincides with $Q^2$ which is a gauge invariant quantity.
If we choose a different gauge such as, for example, the longitudinal one, 
the Mukhanov variable is no longer equal to the scalar field fluctuation 
$\varphi$ and, while $\langle Q^2 \rangle$ takes the same value as in the 
UCG, $\langle \varphi^2 \rangle$ takes a different value.
More generally for any quantity, in a given gauge, one can
always choose a gauge invariant quantity which coincides with it
in that particular gauge. The difficulty is that, except for some
particular cases (such as $\varphi^2$ in the UCG), it may be difficult to
understand the physical meaning of a general gauge invariant quantity. 
This appeares to be the case for the EMT, which is considered in the
next section. A general gauge invariant variable connected with
it in the UCG may not coincide with the value
of the EMT in another gauge. So for the EMT the problem
of the gauge dependence of its value does not appear to be solved.


\section{Approaches to the back-reaction}\label{six}
One may follow different approaches in order to study the backreaction
effects due to cosmological fluctuations.
The main two methods in order to tackle this issue can be described in the
following way.

The first consists of considering only first order perturbations,
then imposing, to first order, the energy and momentum constraints
and finally defining an effective EMT by including all the quadratic terms
present in the Einstein equations. Subsequently one averages everything
over the quantum vacuum and first order quantities disappear.
One finds that the effective EMT which appears in the averaged
Einstein equations is modified by the back-reaction.

The second approach is related to the standard perturbation analysis of
the Einstein equations up to second order. In this case we impose the
energy and momentum constraints and study the inflaton equation of
motion perturbatively up to second order.
One does not define any modified EMT but directly studies
any observable averaged over the quantum vacuum in this
framework.


\subsection{The Energy-Momentum Tensor of Cosmological Fluctuations}
\label{eight1}

On following the first approach the effective EMT of cosmological fluctuations
is given by
\be
\tau^\mu_\nu \equiv T^{\mu \, {\rm quadratic}}_\nu - M_{\rm pl}^2
G^{\mu \, {\rm quadratic}}_\nu \, .
\ee
This method of considering the EMT of gravitational fluctuations is
treated in textbooks \cite{weinberg} and has also been
previously used in \cite{ABML,ABMP,AF,FMVV69}.
In this scheme one considers the modified Einstein equations
$M_{\rm pl}^2 G^{\mu\, (0)}_\nu = T^{\mu\, (0)}_\nu +
\langle \tau^\mu_\nu \rangle$
which therefore includes back-reaction effects.

For a generic potential, with an effective mass $V_{\phi\phi}$ different from
zero, the leading terms in the slow-roll parameter of the energy density are
(see \cite{FMVV69}):
\begin{eqnarray}
\langle \tau^0_0 \rangle & \sim & -\frac{V_{\phi \phi}}{2}
\langle \varphi^2 \rangle - 6 \dot H \langle \varphi^2 \rangle
\nonumber \\
& \equiv & - \varepsilon
\sim - \frac{V_{\phi \phi}}{2} \langle \varphi^2 \rangle \left( 1
- 4 \frac{\epsilon}{\eta}
\right) \,.
\label{tauzerozero}
\end{eqnarray}

Analogously the average pressure is:
\be
\langle \tau^i_j \rangle \equiv p \, \delta^i_j \sim \delta^i_j \left(
- \frac{V_{\phi \phi}}{2} \langle \varphi^2 \rangle -6
\dot H \langle \varphi^2 \rangle
\right) \sim - \varepsilon \, \delta^i_j
\ee
We now restrict the analysis to a power-law potential where $\eta=2 \epsilon$.

The leading behaviour of the renormalized correlator is given by
the infrared part of $\langle\varphi^2\rangle$ (the
adiabatic part does not contribute to this leading value) so on using
Eq.(\ref{tauzerozero}) we can obtain, by Eq.(\ref{phiRENleading}), the leading
value of the energy,  without repeating the adiabatic subtraction, as
\be
\varepsilon_{\rm REN} \sim  -\frac{3}{8 \pi^2} H^2 H_i^2  \,.
\label{EnergyRENleading}
\ee
Such a value leads to a negative contribution in the right hand side
of the Einstein equation (\ref{EE00}) of the order of
$H^2 H_i^2/M_{\rm pl }^2$,
so we have that the average expansion rate appears to be decreased by the
back-reaction of cosmological fluctuations. The importance of back-reaction
is therefore related to the ratio $H_i^2/M_{\rm pl }^2$, if inflation starts
at a Planckian energy then back-reaction during inflation cannot be neglected
as expected. We feel, however, that it is important to underline that for
energy near to the Planck value the semi-classical approach to the quantum
gravity may be questionable.

We want to stress that, in this UCG, the back-reaction effects for a model
of power-law inflation are very differents from the effects for a chaotic
model $\frac{m^2}{2} \phi^2$ found in \cite{FMVV69} and given by a negative
effective energy  of the order of $H_i^6/H^2$. While in the case under
consideration the effective energy has the same time behaviour as the
background energy $3  M_{\rm pl }^2 H^2$, in the chaotic case it grows
with respect to the background value and a negligible back-reaction at the
beginning of inflation can become important before the end of inflation
(see \cite{FMVV69}).
This is in agreament with what Abramo and Woodard state in \cite{AW2}
on using the longitudinal gauge for the calculation of the backreaction
in a power-law model of inflation, namely that the shape of the inflation
potential can have an enormous impact on back-reaction.

For the sake of completenes we have to repeat that the problem of the gauge 
dependence of this effective energy as yet does not appear to be solved.

\subsection{Back-Reaction on the Geometry}\label{eight2}
In the perturbative approach to the Einstein equations any
back-reaction effect is analyzed by evaluating perturbatively
quantities which characterize the geometry (as, for example, the
expansion scalar $\Theta$). As stated in Sec. \ref{five} the
leading effect in the renormalized quantities, for models with an
effective mass different from zero, is given by the infrared part
of the non renormalized integrals, so we shall use the
long-wavelength approximation in the treatment of the second order
scalar perturbations (see \cite{FMVV69,SOGIP}). In the
long-wavelength limit, from the first order equation of motion
Eqs.(\ref{Eq_mukhanov}) and (\ref{Eq_resc_mukhanov}), one obtains
$\varphi \simeq C \frac{\dot{\phi}}{H}$, where $C$ is a constant
in time, thus in this limit the non-local spatial contribution
$\tilde{s}$ (Eq.(\ref{nonlocal})) has an ordinary behavior given,
for the isotropic case, by $\varphi \dot{\varphi}/(4 M _{pl}^2
H)$. In general on using this limit on the right hand side of
Eq.(\ref{Eq_mukhanov_second}) one obtains the following result
(valid to all orders in the slow-roll parameters):
\begin{eqnarray}
 & & \varphi^2 \left\{ \frac{1}{M_{\rm pl }^2} \left[\frac{1}{2}
 \frac{d^2}{dt^2} \left(\frac{\dot{\phi}}{H}\right)+ \frac{3}{4}H
\frac{d}{dt}\left(\frac{\dot{\phi}}{H}\right)+\frac{1}{2}\frac{H}{\dot{\phi}}
\left(\frac{d}{dt}\left(\frac{\dot{\phi}}{H}\right)\right)^{2}\right]
\right. \nonumber \\ & & \left. +
 \frac{H}{2 \dot{\phi}^2}\frac{d^3}{dt^3}\left(\frac{\dot{\phi}}{H}\right)+\frac{3}{2}
\frac{H^2}{\dot{\phi}^2}
\frac{d^2}{dt^2}\left(\frac{\dot{\phi}}{H}\right)- \frac{H^2}{2
\dot{\phi}^3}\frac{d}{dt}
\left(\frac{\dot{\phi}}{H}\right)\frac{d^2}{dt^2}\left(\frac{\dot{\phi}}{H}
\right) -\frac{3}{2}\frac{H^3}{\dot{\phi}^3}
\left[\frac{d}{dt}\left(\frac{\dot{\phi}}{H} \right)\right]^{2}
\right\} \,. \label{Membro_destro}
\end{eqnarray}

For the case under consideration, namely power-law inflation,
$\frac{\dot{\phi}}{H}$ is a constant and, consequently, the
inhomogeneous term of Eq.(\ref{Eq_mukhanov_second}) is equal to
zero in the long-wavelength limit. A coupling between the modes
of different orders in perturbation theory could then occur only
in the ultraviolet. Thus, in the long-wavelength approximation, the
equation of motion for the second order field fluctuation
$\varphi^{(2)}$ is the same as that for the first order one, and one can
write in the infrared an equation for a total fluctuation
$\tilde{\varphi}=\varphi+\varphi^{(2)}$ which is to be constrained by the 
initial condition. 
This is equivalent to saying that since the first and the second
order scalar perturbations $\varphi$ and $\varphi^{(2)}$ are
decoupled, the second order contribution does not give us any new
information, with respect to the first order, and it can be seen
as a renormalization of the first order one. This is the only
useful information which can be gained by the perturbative
approach. 
We feel that the significance of the above considerations and results
deserves further investigations.


\section{Discussion and Conclusions}\label{seven}
The renormalized EMT of cosmological fluctuations for
a power-law model of inflation has been studied. As in our
previous work
\cite{FMVV69} we have
self-consistently taken into account the gravitational fluctuations
accompanying the scalar field fluctuations in a UCG.

We find that the renormalized EMT of cosmological fluctuations
during slow-rollover carries
negative energy density (due to the inclusion of gravitational fluctuations
\cite{FMVV69}) with a de Sitter like equation of state to
leading order. The average expansion rate appears to be decreased by the
back-reaction of cosmological fluctuations and the leading value of the
effective energy is of the order of $H^2H_i^2$, so the back-reaction effects
cannot be neglected if inflation started at Planckian energies, i.e., at
$H_i\sim M_{\rm pl}$.

We stress that the back-reaction effects for this model
are very differents from the effects for a chaotic
model $\frac{m^2}{2} \phi^2$ (\cite{FMVV69}) given by a negative
effective energy  of the order of $H_i^6/H^2$. While in the case under
consideration the effective energy has the same time behaviour as the
background energy $3  M_{\rm pl }^2 H^2$, in the chaotic case it grows with
respect to the background value and also a negligible back-reaction at the
beginning of inflation can become important before the end of inflation
(see \cite{FMVV69}). The back-reaction effects in inflation are strongly
dependent on the shape of the potential (see \cite{AW2} for similar
consideration in longitudinal gauge).

In this model of power-law inflation and in the long-wavelength
approximation we also find that the first and second order scalar
perturbations $\varphi$ and $\varphi^{(2)}$ are decoupled and they
have the same equation of motion, the second order can then be
seen just as a renormalization of the first order with no other
physical information.

To conclude we find in the Appendix the fourth order adiabatic
expansion for a general potential
$V(\phi)$. Let us note that to obtain this general fourth order adiabatic
expansion  a crucial point is
to distinguish between the different adiabatic orders, the effective mass
$V_{\phi \phi}$ obtained by a double field derivative is of adiabatic order 0
while the quantities obtained by deriving n-times with respect to the time
are of adiabatic order n, independently of their value.


\vspace{0.5cm}
\centerline{\bf Acknowledgments}
\vspace{0.2cm}

I wish to thank Fabio Finelli, Gian Paolo Vacca and Giovanni Venturi for
useful discussions and comments on the manuscript.


\section{Appendix: The fourth order adiabatic expansion}
In order to remove the divergences which appear in the integrated
quantities as poles in the $\Gamma$ functions, we shall employ the method
of {\em adiabatic subtraction} \cite{adiabaticZS, adiabatic}.
Such a method consists in
replacing our functions with an expansion in powers of derivatives of the
logarithm of the scale factor.
This expansion
coincides with the adiabatic expansion introduced by Lewis in \cite{lewis}
for a time dependent oscillator.

Usually it is more convenient to formulate the adiabatic expansion by
using the modulus of mode functions $x_k =
|\varphi_{\bf k}/\sqrt{2}|$ and
the conformal time $\eta$ \cite{adiabatic} ($d \eta = d t/a $).
We follow this procedure and write an expansion in derivatives with
respect to the conformal time (denoted by $'$) for $x_k$. We then go back
to cosmic time and insert the expansion in the expectation values
we wish to compute.  Adiabatic expansions in cosmic time and
conformal time lead to equivalent results, because of
the explicit covariance under time reparametrization \cite{HMPM}.

The variable $x_k$ satisfies
the following Pinney equation:

\be
\ddot{x}_k+3 H \dot{x}_k+
\left [\frac{k^2}{a^2}
 V_{\phi \phi} + 2 \left(3 H +
\frac{\dot H}{H}\right)^.
\right ] x_k=\frac{1}{a^6 x^3_k} \,.
\label{EQMPF_k_x}
\ee

Following \cite{FMVV65,FMVV69} we rewrite Eq. (\ref{EQMPF_k_x}) in conformal
time in the following
way:
\be
(a x_k)'' + \Omega_k^2 \, (a  x_k) =
\frac{1}{(a x_k)^3}
\label{conformal_eq_x_k}
\ee
where
\be
\Omega_k^2 = k^2 +a^2 V_{\phi \phi}- \frac{1}{6} a^2 \tilde{R}
\label{conformal_freq_x_k}
\ee
and $\tilde{R}$ is:
\be
\tilde{R} = R - 6 \left(-4 \frac{a^{'2}}{a^4}-2 \frac{a^{''2}}{a^2 a^{'2}}
+ 2 \frac{a^{'''}}{a^2 a^{'}}\right) \,,
\label{ricci_general}
\ee
with $R=6\frac{a''}{a^3}$ the Ricci curvature.
The key point for this calculation is that $V_{\phi \phi}$ is of adiabatic
order 0, while $\tilde{R}$ is of adiabatic order 2, independently of their
value.
From Eqs.
(\ref{conformal_eq_x_k},\ref{conformal_freq_x_k},\ref{ricci_general})
one obtains the expansion for $x_k$ up to the fourth adiabatic order:
\be
x_{k}^{(4)}=\frac{1}{a}\frac{1}{\Omega_k^{1/2}}
\left( 1-\frac{1}{4} \epsilon_2+\frac{5}{32}\epsilon_2^2-
\frac{1}{4}\epsilon_4 \right)
\label{fourth_1}
\ee
where $\Omega_k$ is defined in Eq. (\ref{conformal_freq_x_k}) and
$\epsilon_2 \,, \epsilon_4$ are given by:
\begin{eqnarray}
\epsilon_2&=&-\frac{1}{2}\frac{\Omega_k^{''}}{\Omega_k^3}+\frac{3}{4}
\frac{\Omega_k^{'2}}{\Omega_k^4} \nonumber \\
\epsilon_4 &=& \frac{1}{4}\frac{\Omega_k^{'}}{\Omega_k^3}\epsilon_2^{'}-
\frac{1}{4}\frac{1}{\Omega_k^2}\epsilon_2^{''}
\end{eqnarray}

The solution in Eq. (\ref{fourth_1}) must be expanded again since $\tilde{R}$
is of adiabatic order 2. Therefore $x_k^{(4)}$, for a general potential, is:
\begin{eqnarray}
x_{k (4)} &=& \frac{1}{c^{1/2}}\frac{1}{\Sigma_k^{1/2}} \Bigl\{
1+\frac{1}{4}c\frac{\tilde{R}}{6}\frac{1}{\Sigma_k^2}+\frac{5}{32}c^2
\frac{\tilde{R}^2}{36} \frac{1}{\Sigma_k^4} \nonumber \\
 & &
+\frac{1}{16}\frac{1}{\Sigma_k^4}\left[c^{''} V_{\phi \phi} +2 c^{'}
V_{\phi \phi}^{'}+c V_{\phi \phi}^{''}-c^{''}
\frac{\tilde{R}}{6}
 -2 c^{'}
 \frac{\tilde{R}^{'}}{6}-c\frac{\tilde{R}^{''}}{6}\right]
\nonumber \\
 & &  - \frac{5}{64}\frac{1}{\Sigma_k^6}
\left[ \left(c^{'} V_{\phi \phi}+c V_{\phi \phi}^{'}\right)^2+2
\left(c^{'} V_{\phi \phi}+c V_{\phi \phi}^{'}\right)\left(
 -c^{'} \frac{\tilde{R}}{6}- c \frac{\tilde{R}^{'}}{6}\right)\right]
\nonumber \\
 & & + \frac{9}{64}\frac{1}{\Sigma_k^6} c \frac{\tilde{R}}{6}
\left(c^{''}V_{\phi \phi}+2 c^{'}
V_{\phi \phi}^{'}+c V_{\phi \phi}^{''}\right) -
\frac{65}{256}  \frac{1}{\Sigma_k^8} c \frac{\tilde{R}}{6}
\left(c^{'}V_{\phi \phi}+c V_{\phi \phi}^{'}\right)^2+
\frac{5}{32}\epsilon_{2*}^2- \frac{1}{4}\epsilon_{4*}   \Bigr\}
\label{EA4}
\end{eqnarray}

where $c=a^2$ and
\begin{eqnarray}
\Sigma_k&=&(k^2 + a^2 V_{\phi \phi})^{1/2} \nonumber \\
\epsilon_{2*}&=&-\frac{1}{2}\frac{\Sigma_k^{''}}{\Sigma_k^3}+\frac{3}{4}
\frac{\Sigma_k^{'2}}{\Sigma_k^4} \nonumber \\
\epsilon_{4*} &=&
\frac{1}{4}\frac{\Sigma_k^{'}}{\Sigma_k^3}\epsilon_{2*}^{'}-
\frac{1}{4}\frac{1}{\Sigma_k^2}\epsilon_{2*}^{''}\,.
\end{eqnarray}


\end{document}